\documentclass[aps,onecolumn,superscriptaddress,nofootinbib]{revtex4-2}
\usepackage{bm}
\usepackage{bbm}
\usepackage{epsfig,graphics,graphicx}
\usepackage{amsmath,amssymb,amsfonts}
\usepackage{color}
\usepackage{epstopdf}
\usepackage{slashed}
\usepackage[
colorlinks=true,
linkcolor=black,
breaklinks=true,
urlcolor=blue,
citecolor=fuchsia]{hyperref}

\usepackage[font=footnotesize, labelfont=bf]{caption,subcaption}
\usepackage{booktabs}
\usepackage{dcolumn}
\usepackage{makecell}
\usepackage{multirow}
\newcolumntype{d}{D{.}{.}{5}}

\definecolor{blueviolet}{rgb}{0.541, 0.169, 0.886}
\definecolor{fuchsia}{rgb}{1.0, 0, 1.0}
\newcommand{\be}{\begin{equation}}
	\newcommand{\ee}{\end{equation}}
\newcommand{\ba}{\begin{eqnarray}}
	\newcommand{\ea}{\end{eqnarray}}

\newcommand*{\Rom}[1]{\uppercase\expandafter{\romannumeral #1\relax}}

\begin{document}
\title{Dispersion-theoretical analysis of the electromagnetic form factors of the $\Lambda$ hyperon}
\author{Yong-Hui Lin}
\affiliation{Helmholtz Institut f\"ur Strahlen- und Kernphysik and Bethe Center
	for Theoretical Physics, Universit\"at Bonn, D-53115 Bonn, Germany}
\author{Hans-Werner Hammer}
\affiliation{Technische Universit\"at Darmstadt, Department of Physics,
	Institut f\"ur Kernphysik,\\ 64289 Darmstadt, Germany}
\affiliation{ExtreMe Matter Institute EMMI and Helmholtz Forschungsakademie Hessen f\"ur
	FAIR (HFHF),\\ GSI Helmholtzzentrum f\"ur Schwerionenforschung GmbH,
	64291 Darmstadt, Germany}
\author{Ulf-G. Mei{\ss}ner}
\affiliation{Helmholtz Institut f\"ur Strahlen- und Kernphysik and Bethe Center
   for Theoretical Physics, Universit\"at Bonn, D-53115 Bonn, Germany}
\affiliation{Institute for Advanced Simulation and Institut f{\"u}r Kernphysik,
            Forschungszentrum J{\"u}lich, D-52425 J{\"u}lich, Germany}
\affiliation{Tbilisi State University, 0186 Tbilisi, Georgia}
\date{\today}
%
\begin{abstract}
  The electromagnetic form factors of the $\Lambda$ hyperon
  in the time-like region are determined
  precisely  through a dispersion-theoretical analysis
  of the world data for the cross section of the annihilation process 
  $e^+e^-\to \bar{\Lambda}{\Lambda}$. The spectral function is represented
  by a superposition of narrow and broad vector meson poles. We test different
  scenarios for the spectral function and obtain a good description of
  the world data in the time-like region. The uncertainties in the
  extracted form factors are estimated by means of the bootstrap sampling
  method. The analytical continuation of the form factors to the
  space-like region introduces large errors due to the lack of data.
  When the electric $\Lambda$ radius from chiral perturbation theory is taken
  as a constraint, the magnetic radius is predicted as
  $r_M = 0.681 \pm 0.002$ fm. We also extract various vector meson to baryon
  coupling constants.
\end{abstract}
\maketitle

\section{Introduction}

The precise determination of the electromagnetic form factors (EMFFs) of the
nucleon has become an urgent task since the emergence of notable ``proton radius
puzzle" in 2010, namely the tension between the muonic determination of proton
charge radius and previous measurements based on the electron-proton scattering
and hydrogen spectroscopy~\cite{Pohl:2010zza,Antognini:2013txn,Antognini:2022xoo},
though it must be said that the dispersion theoretical analysis of the
scattering data always led to a value consistent with the one from muonic hydrogen,
see the discussion in Ref.~\cite{Lin:2021umz}.
The nucleon EMFFs are accessed experimentally through elastic
electron-proton scattering $e^-p\to e^-p$ as well as electron scattering
on light nuclei (as no neutron target exists) for the space-like region, and
the $\bar{p}p$ annihilation process $\bar{p}p\to e^+e^-$ and likewise the
reactions $e^+e^-\to \bar{p}p$ and $e^+e^-\to \bar{n}n$ for the time-like
region (see, e.g., Refs.~\cite{Denig:2012by,Pacetti:2014jai,Punjabi:2015bba}
for recent reviews).  In this context, some impressive experimental
results have been reported over the last decade. The PRad collaboration
measured the differential cross sections of $e^- p$ scattering
down to an unprecedented momentum transfer of
$Q^2=2.1\times 10^{-4}$~GeV$^2$~\cite{Xiong:2019umf} while 
the BaBar~\cite{BaBar:2013ves} and
BES\Rom{3}~\cite{BESIII:2015axk,BESIII:2019tgo,BESIII:2019hdp,BESIII:2021rqk}
collaborations measured the total cross sections for $e^+e^-\to \bar{p}p$
over a large range of center-of-mass energies. And still many experimental and
theoretical efforts continue.
On the theoretical side, we mention in particular a high-precision dispersion-theoretical analysis
of the world data of the nucleon EMFFs in the space- and time-like regions
that  was reported in Ref.~\cite{Lin:2021xrc} last year.
The statistical uncertainties of the extracted form factors were estimated
using the bootstrap method, while systematic errors were determined from
variations of the spectral functions. This work further solidified the
earlier findings of a small proton charge radius $r_p = 0.84\,$fm with subpercent
accuracy.

There has also been increasing interest in the electromagnetic structure of 
hyperons in the past two decades, both from the
experimental~\cite{DM2:1990tut,BaBar:2007fsu,Dobbs:2014ifa,Dobbs:2017hyd,
Belle:2017caf,BESIII:2017hyw,BESIII:2019nep,BESIII:2020uqk,BESIII:2019cuv,
BESIII:2020ktn,BESIII:2021ccp}
and theoretical~\cite{Baldini:2007qg,Dalkarov:2009yf,Faldt:2013gka,
Haidenbauer:2016won,Cao:2018kos,Yang:2017hao,Yang:2019mzq,Xiao:2019qhl,
Dubnicka:2018,Faldt:2017kgy,Perotti:2018wxm,Ferroli:2020xnv,Ramalho:2019koj,
Haidenbauer:2020wyp,Li:2021lvs,Bystritskiy:2022nqn} side.
Compared to the nucleon, the development of experiments exploring
the hyperon EMFFs is somewhat lagging behind since the hyperons are unstable.
Therefore hyperon targets for elastic electron scattering experiments that 
access the EMFFs of hyperons in the spacelike region are not available.
The main source of information are measurements of the reaction
$e^+e^-\to \bar{\Lambda}\Lambda$ that depends on the EMFFs in the
time-like region.
For a recent review of the current experimental status on $\Lambda$ EMFFs
see Ref.~\cite{Zhou:2022jwr}. A recent improvement of the data base
is  provided by the cross section for $e^+e^-\to \bar{\Lambda}\Lambda$
around $4$~GeV reported by the BES\Rom{3} Collaboration
last year~\cite{BESIII:2021ccp}.
It enriches the data base of the $\Lambda$ EMFFs, which previously only covered
the near-threshold region, to a great extent.

To date, only a subset of these $\bar{\Lambda}\Lambda$ data has been studied
theoretically~\cite{Haidenbauer:2016won,Cao:2018kos,Yang:2019mzq,
Haidenbauer:2020wyp,Li:2021lvs,Bystritskiy:2022nqn}. 
In particular, the previous near-threshold
data were analyzed in Refs.~\cite{Haidenbauer:2016won,Haidenbauer:2020wyp},
which used phenomenological $\bar{Y}Y$ potential models, that are constrained
by final-state interaction (FSI) effects in reactions like
$\bar{p}p\to \bar{Y}Y$, to calculate the EMFFs of hyperons.
Moreover, Ref.~\cite{Cao:2018kos} utilized Fano-type form factors, 
that include the interference between several resonances and 
one continuum background constructed based on
perturbative QCD (pQCD) predictions, 
to fit those data near threshold. This analysis found that
the excited $\phi$ meson $\phi(2170)$ is required to reproduce 
the close-to-threshold enhancement observed in the
$e^+e^-\to \bar{\Lambda}\Lambda$ reactions. Similar interpretations 
were also investigated in Refs.~\cite{Yang:2019mzq,Li:2021lvs} using 
the vector meson dominance (VMD) parameterization for the EMFFs 
of the $\Lambda$ hyperon. In contrast, the work of
Ref.~\cite{Bystritskiy:2022nqn} only focused on the newest measurements
around $4$~GeV by the BES\Rom{3} group. It was found that the dip near the
mass of $\psi(3770)$ in the cross sections of $e^+e^-\to\bar{\Lambda}\Lambda$ 
gets contributions from both $D$-meson loops and the three
gluon charmonium annihilation mechanism.

The present work seeks to obtain a fit to the full data basis of
$\bar{\Lambda}\Lambda$ production data collected so far based on dispersion
theory~\cite{Chew:1958zjr,Federbush:1958zz,Hohler:1976ax,Mergell:1995bf}.
The spectral function of the EMFFs is parametrized in terms of
narrow and broad vector meson poles and includes
the constraints from unitarity, analyticity and crossing
symmetry. Moreover, it is consistent with the strictures
from pQCD at very large momentum transfer \cite{Lepage:1980fj}.
The uncertainties in the extracted form factors are estimated by means of
the bootstrap sampling method. We already note here that for an estimation
of the systematic uncertainty, which can be accessed by varying the
number of poles in the spectral function~\cite{Hohler:1976ax,Lin:2021umz},
the data base is simply too sparse. A recent review of the dispersion-theoretical
formalism is given in Ref.~\cite{Lin:2021umz}.
Here we extend this dispersive strategy to the hyperon
cases in a straightforward way.

The paper is organized as follows:
in Section~\ref{sec.Born} the total cross section of the process
$e^+e^- \to \Lambda\bar{\Lambda}$ in Born approximation is presented and
the details of the dispersion-theoretical parameterization of the
electromagnetic form factors of $\Lambda$ hyperon as well as
the fit strategy are illustrated.
Section~\ref{sec.Results} contains the numerical results
for  the EMFFs of the $\Lambda$ hyperon and their discussion.
A summary and conclusion is given in Section~\ref{sec.Summary}.

\section{Formalism}\label{sec.Born}
We briefly introduce the basic formulae for the analysis of the EMFFs of the
$\Lambda$ hyperon in the dispersion-theoretical framework. When assuming
one-photon exchange as the sole contribution, the so-called Born
approximation, the total cross section of the annihilation reaction $e^+e^-\to \bar{Y}Y$
can be written as~\cite{Haidenbauer:2020wyp},
\begin{equation}\label{eq:xs}
	\sigma_{e^+e^-\to \bar{Y}Y}=\frac{4\pi\alpha^2\beta}{3s}C(s)
	\left[|G_M(s)|^2+\frac{2m_Y^2}{s}|G_E(s)|^2\right],
\end{equation}
where $Y=\Lambda$, $\Sigma$, $\Xi$ denotes the hyperons with
$\bar{Y}$ the corresponding anti-hyperons. Moreover,
$\alpha\approx1/137.036$ is the
fine-structure constant and $\beta=k_Y/k_e$ denotes a phase-space factor.
Here, $k_Y$ and $k_e$ are the moduli of the center-of-mass three-momenta
in the outgoing $\bar{Y}Y$ and incoming $e^+e^-$ systems, satisfying
$s=4(m_Y^2+k_Y^2)=4(m_e^2+k_e^2)$, with $\sqrt{s}$ the total energy
and $m_Y$ ($m_e$) the hyperon (electron) masses. $C(s)$ represents the
S-wave Sommerfeld-Gamow factor defined by $C=y/(1-e^{-y})$ with
$y=\pi\alpha m_Y/k_Y$. Note that $C(s)\equiv 1$ for the neutral hyperons
($\Lambda$, $\Sigma^0$, $\Xi^0$). The complex functions $G_E(s)$ and $G_M(s)$
are the Sachs electric and magnetic form factors of the hyperons
in the time-like region. They are accessible in this reaction for
$s\geq 4m_Y^2$. When the functions are analytically continued to negative values
of $s$, i.e. to the space-like region, they are real and describe the
elastic scattering of electrons of the hyperons. The variable
$-s\equiv Q^2$ then specifies the four-momentum transfer $Q^2$ in
the elastic scattering process.

Since the separation of $G_E$ and $G_M$ requires angular distributions,
the time-like experimental data is often given in the form
of the effective form factor $G_{\rm eff}$, which is defined by
\begin{equation}\label{eq:geff}
	|G_{\rm eff}(s)|=\sqrt{\frac{\sigma_{e^+e^-\to \bar{Y}Y}(s)}
	{\frac{4\pi\alpha^2\beta}{3s}C(s)\left(1+\frac{2m_Y^2}{s}\right)}}.
\end{equation}

When employing the dispersion-theoretical analysis,
it is convenient to express the Sachs form factors in terms of the Dirac
($F_1$) and Pauli ($F_2$) form factors,
\begin{equation}
	G_M=F_1+F_2,\quad G_E=F_1+\frac{s}{4m_Y^2}F_2,
\end{equation}
with the normalization at zero momentum transfer given by
$F_1(0)=G_E(0)=0$ and $F_2(0)=G_M(0)=\mu_{\Lambda}$ for the
electrically neutral $\Lambda$ hyperon. Here
$\mu_\Lambda=-0.613\hat{\mu}_N=-0.723\hat{\mu}_\Lambda$
is the magnetic moment of the $\Lambda$ hyperon~\cite{Workman:2022ynf}
with  $\hat{\mu}_N\equiv e/(2m_N)$ and $\hat{\mu}_\Lambda\equiv e/(2m_\Lambda)$.

The strategy of the dispersion-theoretical
analysis for the EMFFs of the $\Lambda$ hyperon is quite similar to the nucleon
case which is explained comprehensively in the review~\cite{Lin:2021umz}. The
dispersion relation for a generic form factor $F(s)$ is written as
\begin{equation}
	F(s)=\lim_{\epsilon\to 0^+}\frac1\pi\int_{s_0}^\infty d s^\prime\frac{{\rm Im}~F(s^\prime)}{s^\prime-s-i\epsilon},
\end{equation}
where ${\rm Im}~F$ in the time-like region, also called the spectral function, 
is required as input. Furthermore,
$s_0$ denotes the threshold of the lowest cut of the form factor
$F(s)$. In the nucleon case,
$s_0=4M_\pi^2~(9M_\pi^2)$ for the isovector (isoscalar) form factors.
For the $\Lambda$ hyperon, there is only an isoscalar
contribution, i.e., $F^\Lambda_i=F_i^s$ with $i=1,2$ and $s_0=9M_\pi^2$,
since the $\Lambda$ is a
isospin singlet. The contribution of the three-pion cut is expected
to be small except for the $\omega$ contribution. See
Ref.~\cite{Bernard:1996cc} for an explicit calculation in the nucleon case,
which shows that the anomalous threshold in the three-pion channel is
effectively masked by the phase space behavior.

Therefore, the spectral functions for the $\Lambda$ isoscalar FFs are
parameterized in terms of effective vector meson poles, without
any explicit continuum contributions.
This leads to the following representation of the FFs:
\begin{align}\label{eq:drffs}
	F_i^s(s)=\sum_{V=\omega,\phi,s_1,\cdots}\frac{a_i^V}{M_V^2-s-i\epsilon}
	+\sum_{V=\phi_{2170},\psi_{3770},S_1,\cdots}\frac{a_i^V}{M_V^2-s-iM_V\Gamma_V}.
\end{align}
Here, the broad vector meson poles are introduced to generate the non-zero
imaginary part of the EMFFs of the
$\Lambda$ in the timelike region. In the narrow
sector, two physical states, the $\omega$ and $\phi$ mesons, are included as in
our previous analyses to the nucleon EMFFs~\cite{Lin:2021umz,Lin:2021xrc}. 
Following the previous studies on the $\bar{\Lambda}\Lambda$ data 
in the near-threshold region~\cite{Cao:2018kos,Yang:2019mzq,Li:2021lvs} 
and works on the newest BES\Rom{3} data set
around $4$~GeV~\cite{BESIII:2021ccp,Bystritskiy:2022nqn}, 
the $\phi(2170)$ and $\psi(3770)$ states are included in the broad sector. 
The PDG values~\cite{Workman:2022ynf} for the masses and widths of
those physical vector states are used as input in our analysis.
The fit parameters are therefore
the various meson residua $a_i^V$ and the masses and widths of the
additional vector mesons $s_i$, $S_i$. Since these parametrize
continuum contributions, we require the widths of the $S_i$ to
be larger than the width of the physical $\psi(3770)$.
Finally, several physical
constraints are included in the parametrizations of Eq.~\eqref{eq:drffs}.
First, two normalization conditions for the values of FFs at zero
momentum transfer must be fulfilled. Second, three constraints from
the superconvergence relations inferred from the pQCD
predictions for the asymptotic behavior of the EMFFs at very large
momentum transfer are taken into account. They are given by                 
\begin{equation}
	\int_{s_0}^\infty {\rm Im}~F_i(s) s^n ds=0, \quad i=1,2,
\end{equation}
with $n=0$ for $F_1$ and $n=0,1$ for $F_2$. Third, the electric charge radius of
the $\Lambda$ hyperon is fixed to the value $\langle r_E^2\rangle=0.11\pm 0.02~{\rm fm}^2$ 
that was calculated in chiral perturbation theory~\cite{Kubis:2000aa}.
This constraint is useful since the energy region covered by the time-like data
starts at $s=4m_\Lambda^2$ and thus the slope at $s=0$
can not be constrained well by them.

\section{Fit Strategy and Results}\label{sec.Results}

The first step is to find the best configuration for the EMFFs of the $\Lambda$
hyperon, i.e., the numbers of the narrow poles $s_i$ and the broad poles $S_i$ in
Eq.~\eqref{eq:drffs}. We start by including only the narrow physical poles
($\omega$ and $\phi$ mesons) with fixed masses while the
$\phi(2170)$ and $\psi(3770)$ states are not included \textit{a priori}.
In the second step, we increase the numbers of
narrow poles $s_i$ and broad poles $S_i$ one by one. All
pole masses except those of the $\omega$ and $\phi$ poles
and all residua are fit parameters. Using
this procedure, the database for the $\bar{\Lambda}\Lambda$ production composed
of the measurements by the DM2~\cite{DM2:1990tut}, BaBar~\cite{BaBar:2007fsu},
CLEO~\cite{Dobbs:2017hyd} and BES\Rom{3}~\cite{BESIII:2017hyw,BESIII:2019nep,BESIII:2021ccp} collaborations
is fitted with $7$ constraints: $2$ normalization conditions, 
$3$ superconvergence relations, and $1$ radius condition
for $\langle r_E^2\rangle$.

\begin{figure}[t!] 
	\includegraphics*[width=0.49\linewidth,angle=0]{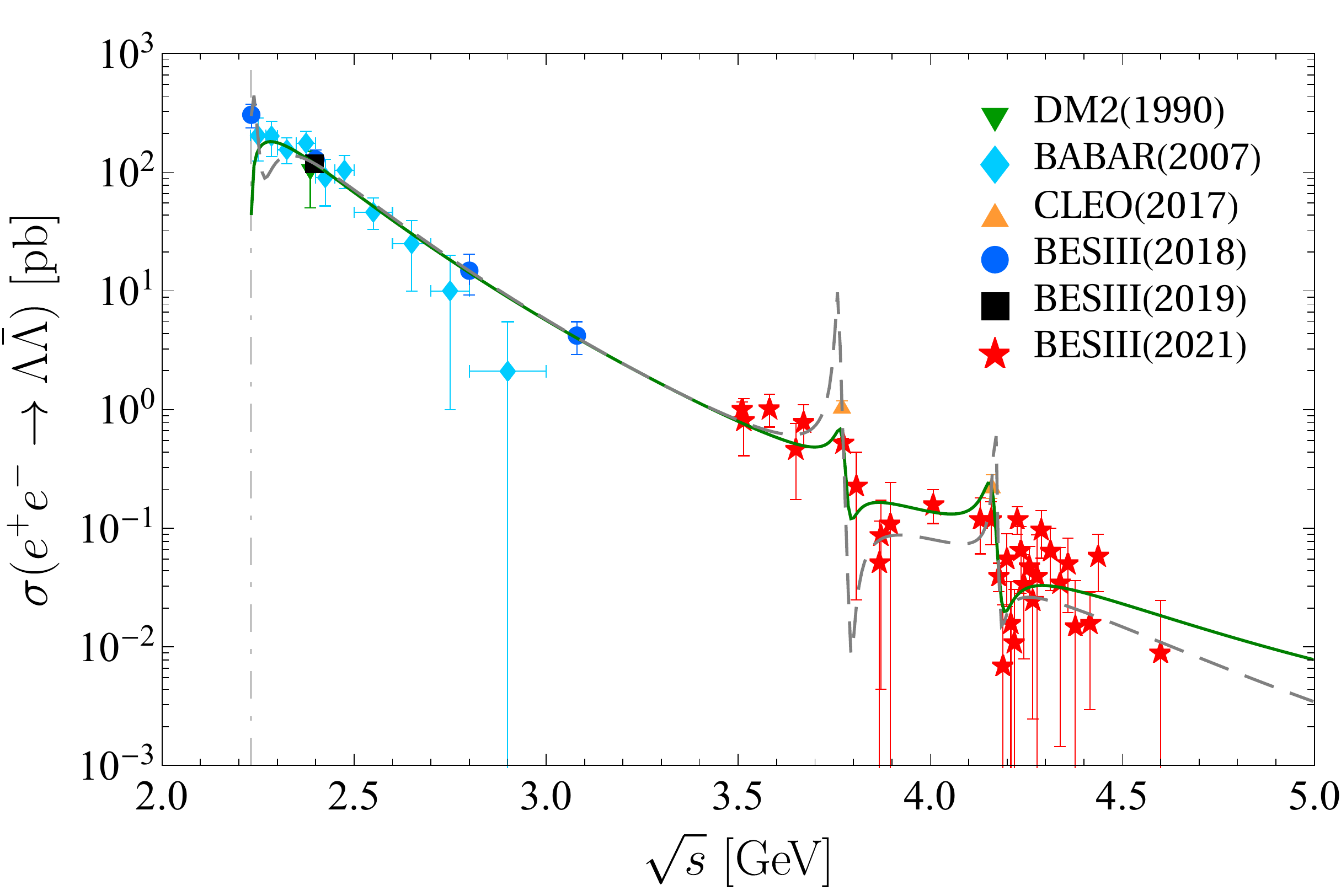}~~~~
	\includegraphics*[width=0.49\linewidth,angle=0]{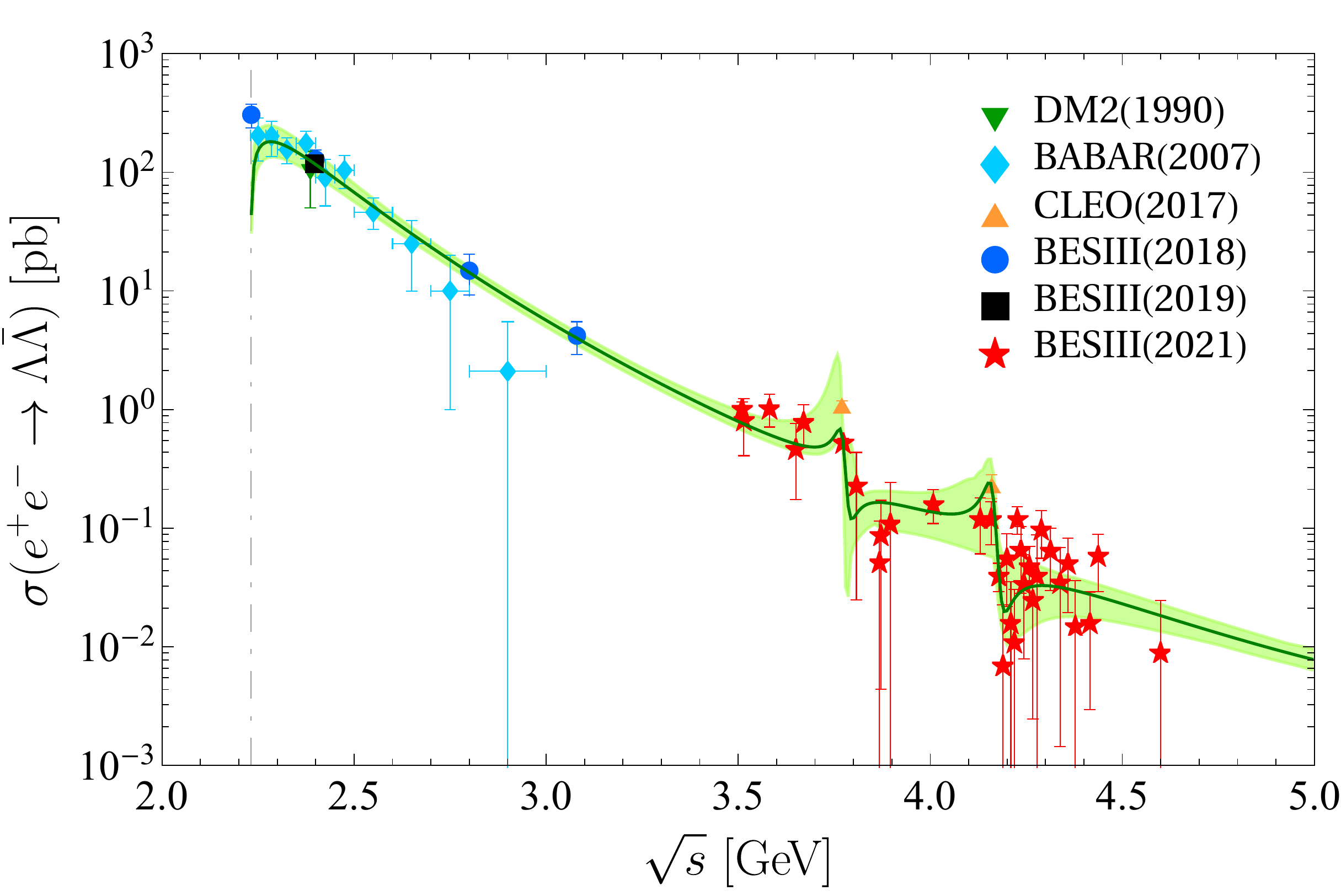} 	
	\caption{Results of the fit to the full data sets for the total
	cross sections of the reaction $e^+e^-\to\bar{\Lambda}\Lambda$ that
	taken from Refs.~\cite{DM2:1990tut,BaBar:2007fsu,Dobbs:2017hyd,BESIII:2017hyw,
		BESIII:2019nep,BESIII:2021ccp}. Left panel: comparison between 
	Fit-\Rom{1} (dashed line) and Fit-\Rom{2} (solid line). 
	Right panel: Fit-\Rom{2} with the error band given by the bootstrap method. 
	The $\bar{\Lambda}\Lambda$ threshold is represented by the vertical dash-dotted lines.
	}
	\label{fig:HEFFs}
	\vspace{-3mm}
\end{figure}

We find that
at least one additional narrow and three broad poles are required for
an acceptable $\chi^2$ in the fit. The reduced $\chi^2$ for the best fit is
$1.531$. In this fit the EMFFs of the $\Lambda$ hyperon get contributions from
6 pole terms: $\omega$,
$\phi$ and $s_1(2014)$ in the narrow part and $S_1(2232)$, $S_2(3760)$ and
$S_3(4170)$ in the broad part. We refer to this fit as
Fit-\Rom{1}. It is displayed as the dashed line in the left panel
of Fig.~\ref{fig:HEFFs}.
Comparing the pole masses to the mass spectra for the light
vector mesons and charmonia ($c\bar{c}$) listed in
PDG~\cite{Workman:2022ynf}, the effective narrow pole $s_1(2014)$
and the broad pole $S_1(2232)$ obtained in the fit are close to the
physical $\phi(2170)$ state
($M=2162\pm 7$~MeV and $\Gamma=100^{+31}_{-23}$~MeV).
Moreover, the $S_2(3760)$ and $S_3(4170)$ poles are quite close to the
charmonium states $\psi(3770)$ ($M=3773.7\pm 0.4$~MeV
and $\Gamma=27.2\pm 1.0$~MeV) and $\psi(4160)$ ($M=4191\pm 5$~MeV and
$\Gamma=70\pm 10$~MeV), respectively.
When the $\phi(2170)$, $\psi(3770)$ and $\psi(4160)$ states are included
as broad poles with fixed masses in the spectral
function, one gets a relatively large reduced $\chi^2$ of $3.877$.
The main contribution to the $\chi^2$
comes from the data set reported in Ref.~\cite{BESIII:2021ccp}
(labeled as BES\Rom{3}(2021) in the following). 
Our analysis suggests that the contribution to the $\chi^2$ of the data
set BES\Rom{3}(2021) is quite sensitive to the parameters of the effective
poles located in the energy region of the data. This is because there
are large statistical fluctuations in this data set, especially in the region
above $4$~GeV, which was also noticed in Ref.~\cite{Bianconi:2022yjq}.

Because of these observations, we focus in the following on the choice of
the $\Lambda$ spectral function
where the
two physical mesons ($\omega$ and $\phi$) and one floating effective pole
$s_1$ are contained in the narrow sector and the $\phi(2170)$ and $\psi(3770)$
states with fixed masses 
together with one additional floating pole $S_1$ make up the broad sector
(the best solution for this choice of spectral function is
labeled as Fit-\Rom{2}).
It is worth mentioning that we increase the weight of the sixth data point 
(the point at energy of $3.7730$~GeV) of the data set BES\Rom{3}(2021) 
by a factor 10 due to its extremely small error.
\begin{table}[htbp]
	\centering
	\renewcommand\arraystretch{1.2}
	\caption{Comparison between the Fit-\Rom{1} and Fit-\Rom{2}. Fit-\Rom{2} is our preferred fit. The $\ddagger$ represents an input quantity that  was fixed to the physical state from PDG~\cite{Workman:2022ynf} (mass and width are constant in the fit). \label{tab:compare}}
	\begin{tabular}{|p{2.0cm}<{\centering}
			|p{1.2cm}<{\centering}p{1.2cm}<{\centering}p{1.5cm}<{\centering}
			|p{2.4cm}<{\centering}p{2.2cm}<{\centering}p{2.2cm}<{\centering}
			|p{2.0cm}<{\centering}|}
		\hline
		\hline
		Conf. & \multicolumn{3}{c}{narrow($M_V$)} & \multicolumn{3}{c}{Broad($M_V+i\Gamma_V$)} & $\chi^2$ \\
		\Xhline{0.2pt}
		Fit-\Rom{1}  & \multirow{2}*{$\omega(782)^\ddagger$} & \multirow{2}*{$\phi(1020)^\ddagger$} & $s_1(2014)$  & $S_1(2232+i 33)$ & $S_2(3760+i 11)$ & $S_3(4170+i 6)$ & $1.531$ \\
		Fit-\Rom{2}  & & & $s_1(1965)$  & ${\phi(2170+i 100)^\ddagger}$ & ${\psi(3770+i 27)}^\ddagger$ & $S_1(4163+i 32)$ & $2.596$ \\
		\hline
		\hline
	\end{tabular}
\end{table}

A comparison between the pole content and results of Fit-\Rom{1} and Fit-\Rom{2} is given in Table~\ref{tab:compare} 
and Fig.~\ref{fig:HEFFs}, respectively. The reduced $\chi^2$ of Fit-\Rom{2} is $2.596$ which is
compatible with the ``Fit~\Rom{1}" implemented by the BES\Rom{3} Collaboration
in their experimental announcement~\cite{BESIII:2021ccp}. The uncertainties are
estimated by using the bootstrap sampling method. As one can see from the right panel of
Fig.~\ref{fig:HEFFs}, all measurements for the reaction
$e^+e^-\to\bar{\Lambda}\Lambda$ can be described well by Fit-\Rom{2}. Note that the
oscillation in the near-threshold region generated by the interference
of $s_1(2011)$ and $S_1(2232)$ in Fit-\Rom{1} is gone. In Fit-\Rom{1}, 
the data point in BES\Rom{3}(2018)~\cite{BESIII:2017hyw}
close to threshold is reproduced by that oscillation in agreement
with the finding of Ref.~\cite{Li:2021lvs}.

We replace this feature with the simpler single-pole contribution from the
physical state
$\phi(2170)$ in Fit-\Rom{2} since such an oscillation can not be confirmed
definitely by the current experiments.
This is in conjunction with the general strategy to use as few poles
as possible to stabilize the ill-posed problem of
reconstructing the spectral function from experimental data
\cite{sabba,SabbaStefanescu:1978hvt}. However,
a further experimental investigation of this issue is required in the future.
Furthermore, a pole located around $4.163$~GeV is suggested by Fit-\Rom{2}
which could be identified with the physical state $\psi(4160)$. Its
mass is, however, outside the mass range for the $\psi(4160)$
quoted in PDG~\cite{Workman:2022ynf}. The physical
information regarding this pole will be more clear if the statistical
fluctuations in BES\Rom{3}(2021) data can be reduced.
In this fit, we use the electric radius of the $\Lambda$ hyperon as input,
$r_E=\sqrt{\langle r_E^2\rangle}={0.332}~{\rm fm}$ from Ref.~~\cite{Kubis:2000aa}
to have some constraint from the space-like region.
The magnetic radius of the $\Lambda$ hyperon derived from Fit-\Rom{2} is then given by
\begin{equation}
	 r_M=0.681\pm 0.002~{\rm fm}~.
\end{equation}
The obtained magnetic radius  is a bit larger than the value in Ref.~\cite{Yang:2019mzq}, $r_M=0.42$~fm and also the
electric radius given there is smaller than the CHPT determination,  $r_E=0.11$~fm.
The fit parameters of Fit-\Rom{2} are presented in Table~\ref{tab:fitpara} together with the bootstrap error bars.

Here, we only focus on the residua of the $\omega$ and $\phi$ mesons. 
Translating into the couplings $\omega\Lambda\Lambda$ and
$\phi\Lambda\Lambda$~\cite{Lin:2021umz}, one obtains
\begin{equation}
	g_1^{\omega\Lambda\Lambda}=18.04_{-0.11}^{+0.13},\quad g_2^{\omega\Lambda\Lambda}=-34.45_{-0.24}^{+0.20}, 
	\quad g_1^{\phi\Lambda\Lambda}=-16.81_{-0.24}^{+0.19},\quad g_2^{\phi\Lambda\Lambda}=18.99_{-0.20}^{+0.25},
        \label{eq:Lcoups}
\end{equation}
corresponding to a tensor-to-vector coupling ratio of about
$-2$ for the $\omega\Lambda\Lambda$ and $-1$ for $\phi\Lambda\Lambda$
case.
The corresponding $\omega$ couplings to the nucleon derived from the residua  in the
dispersion analysis of nucleon form factors~\cite{Lin:2021umz}
including the statistical (first error) and systematic (second error)
uncertainties
are $g_1^{\omega NN}=18.6\pm 2.0\pm 3.8$ and
$g_2^{\omega NN}=8.4\pm 3.2\pm 5.8$ such that the tensor-to-vector
coupling ratio, $\kappa_\omega= 0.42^{+0.44}_{-1.24}$~\cite{Lin:2021xrc}. It thus that has a
large uncertainty, but is still compatible with zero. Also,
the statistical and systematic errors are added in quadrature when calculating the ratio.
Note that for the nucleon no $\phi$ couplings were given since the
separation of the $\phi$ from the $K\bar{K}$ continuum is ambiguous.

Next we investigate the role of SU(3) flavor symmetry for the vector
couplings. The ratios between the nucleon and $\Lambda$
hyperon vector couplings
are given by $g_1^{\omega \Lambda\Lambda}/g_1^{\omega NN}=1.00_{-0.39}^{+0.65}$ 
and $g_1^{\phi \Lambda\Lambda}/g_1^{\omega NN}=-0.94_{-1.09}^{+0.38}$
which are expressed as $(2/3)\times(5\alpha_{\rm BBV}-2)/(4\alpha_{\rm BBV}-1)$
and $(-\sqrt{2}/3)\times(2\alpha_{\rm BBV}+1)/(4\alpha_{\rm BBV}-1)$,
respectively, with the SU(3) relations~\cite{Ronchen:2012eg}. The ratio
$g_1^{\omega \Lambda\Lambda}/g_1^{\omega NN}$ follows the SU(3) expectation
when $\alpha_{\rm BBV}\geq 0.82$, while the matching condition for 
$g_1^{\phi \Lambda\Lambda}/g_1^{\omega NN}$ is $\alpha_{\rm BBV}\leq 0.8$.
Only the former case can be compatible with the SU(3) symmetry expectation that $\alpha_{\rm BBV}\approx 1.0$.
For the tensor coupling, $g_2^{\omega NN}$ is expected to be zero in SU(3) symmetry and $g_2^{\omega\Lambda\Lambda}$
can be expressed in terms of $g_2^{\omega NN}$ and $g_2^{\rho NN}$. However, the latter quantity  is not available
in the previous dispersion-theoretical analysis of NFFs as the $\rho$ is part of the two-pion continuum.
Further, we can extract the couplings for the $\psi(3770)$. We find $g_1^{\psi(3770)\Lambda\Lambda}=-0.0040\pm 0.0008$
and $g_2^{\psi(3770)\Lambda\Lambda}=0.0012\pm 0.0004$. A similar extraction of the $\phi(2170)$ couplings is not
possible since there is no experimental data for the partial width of $\phi(2170)\to e^+e^-$. 

\begin{table}[htbp]
	\centering
	\renewcommand\arraystretch{1.5}
	\caption{The parameters corresponding to our best fit (Fit-\Rom{2}) together with the
	 bootstrap errors. Masses $M_V$ and widths $\Gamma_V$ are given 
	 in units of GeV and residua $a_i^V$ in $\rm GeV^2$. 
	 The $\ddagger$ indicates that the corresponding parameter was
         taken as input and is not fitted. \label{tab:fitpara}}
	\begin{tabular}{|p{2.0cm}<{\centering}p{2.4cm}<{\centering}p{2.4cm}<{\centering}p{2.4cm}<{\centering}|}
		\hline
		\hline
		Narrow & $\omega(782)$ & $\phi (1020)$ & $s_1$ \\
		\hline
		$M_{\rm narrow}$  & ${0.783}^\ddagger$  & ${1.019}^\ddagger$	&	${1.9647}_{-0.0316}^{+0.0274}$ \\
		$a_1$&	${0.6699}_{-0.0040}^{+0.0050}$	&	${-1.3025}_{-0.0188}^{+0.0151}$	&	${0.6326}_{-0.0110}^{+0.0138}$	\\
		$a_2$  &	${-1.2793}_{-0.0089}^{+0.0074}$  &	${1.4721}_{-0.0155}^{+0.0191}$	&	${-0.1928}_{-0.0102}^{+0.0082}$ \\
		\hline
		\hline
		Broad & $\phi(2170)$ & $\psi(3770)$ & $S_1$ \\
		\hline
		{$M_{\rm broad}$}  & ${2.162}^\ddagger$  & ${3.7737}^\ddagger$	&	${4.1630}_{-0.0252}^{+0.0048}$ \\
		$\Gamma_{\rm broad}$  & ${0.100}^\ddagger$  & ${0.0272}^\ddagger$	&	${0.0321}_{-0.0021}^{+0.0324}$ \\
		$a_1$  & ${-0.0094}_{-0.0035}^{+0.0043}$	&	${-0.0010}_{-0.0002}^{+0.0002}$	&	${-0.0012}_{-0.0024}^{+0.0003}$	\\
		$a_2$  &	${-0.0196}_{-0.0053}^{+0.0032}$  &	${0.0003}_{-0.0001}^{+0.0001}$	&	${0.0005}_{-0.0002}^{+0.0012}$ \\
		\hline
		\hline
	\end{tabular}
\end{table}

\section{Summary}\label{sec.Summary}
In this work, we have analyzed the full set of cross section full data for the reaction $e^+e^-\to\bar{\Lambda}\Lambda$,
including the recent measurements around $s=4$~GeV by BES\Rom{3}
in dispersion theory. The extracted EM form factors of the $\Lambda$ hyperon
from our best fit (Fit-\Rom{2}, see Table~\ref{tab:fitpara}
and right panel of Fig.~\ref{fig:HEFFs})
can describe the world data base with a reduced $\chi^2$ of $2.596$.
The $\Lambda$ spectral function of the best fit contains
two physical mesons, $\omega(782)$ and $\phi(1020)$,
and one floating effective pole
$s_1$ in the narrow sector and the $\phi(2170)$ and $\psi(3770)$
states with fixed masses 
together with one additional floating pole $S_1$ in the broad sector.
A slightly better $\chi^2$ of $1.531$ was obtained in  Fit-\Rom{1}
at the expense of a double
pole structure close to threshold whose experimental status is unclear.
The uncertainties in the extracted form factors are given by means of the
bootstrap approach. An estimate of the systematic errors from a variation
of the number of effective poles is precluded by the sparse data set.
The form factors in the space-like region are only weakly constrained
by the data in the time-like region. Including the value for electric
radius from  the chiral perturbation theory calculation
of Ref.~\cite{Kubis:2000aa} as a constraint, however, a magnetic radius
$\langle r_M^2 \rangle^{1/2}= 0.681\pm 0.002~{\rm fm}$ was extracted.
From the fit, we could determine the vector and tensor couplings of the
$\omega$ and the $\phi$ to the $\Lambda$ hyperon, see Eq.~\eqref{eq:Lcoups}.
We have further extracted the $\omega NN$ couplings and confirm that the
tensor coupling is suppressed and compatible with zero. It is also small
for the $\psi(3770)\Lambda\Lambda$ coupling, where the tensor-to-vector
coupling ratio is $\kappa^{\psi(3770)} = -0.3\pm0.5$.

\acknowledgments
The work of UGM and YHL is supported in
part by  the DFG (Project number 196253076 - TRR 110)
and the NSFC (Grant No. 11621131001) through the funds provided
to the Sino-German CRC 110 ``Symmetries and the Emergence of
Structure in QCD",  by the Chinese Academy of Sciences (CAS) through a President's
International Fellowship Initiative (PIFI) (Grant No. 2018DM0034), by the VolkswagenStiftung
(Grant No. 93562), and by the EU Horizon 2020 research and innovation programme, STRONG-2020 project
under grant agreement No 824093. HWH was supported by the Deutsche Forschungsgemeinschaft (DFG, German
Research Foundation) -- Projektnummer 279384907 -- CRC 1245
and by the German Federal Ministry of Education and Research (BMBF) (Grant
no. 05P21RDFNB).

\end{document}